\documentclass[%
reprint,
superscriptaddress,
 amsmath,amssymb,
 aps,]{revtex4-2}
\usepackage{graphicx}
\usepackage{dcolumn}
\usepackage{bm}
\usepackage{appendix}
\usepackage{soul}
\usepackage{physics}
\usepackage{float}
\usepackage{multirow}
\usepackage{hyperref}
\hypersetup{
    colorlinks=true,
    linkcolor=blue,
    filecolor=magenta,      
    urlcolor=blue,
    citecolor = blue,
}

\begin{document}

\title{Identification of particle mixtures using machine-learning-assisted laser diffraction analysis}

\author{Arturo Villegas}
\affiliation{ICFO—Institut de Ciencies Fotoniques, the Barcelona Institute of Science and Technology, 08860 Castelldefels (Barcelona), Spain}
\email{arturo.villegas@icfo.eu}

\author{Mario A. Quiroz-Ju\'{a}rez}
\affiliation{Instituto de Ciencias Nucleares, Universidad Nacional Aut\'onoma de M\'exico, Apartado Postal 70-543, 04510 Cd. Mx., M\'exico}
\affiliation{Departamento de F\'{i}sica, Universidad Aut\'onoma Metropolitana Unidad Iztapalapa, San Rafael Atlixco 186, 09340 Cd. Mx., M\'exico}

\author{Alfred B. U'Ren}
\affiliation{Instituto de Ciencias Nucleares, Universidad Nacional Aut\'onoma de M\'exico, Apartado Postal 70-543, 04510 Cd. Mx., M\'exico}

\author{Juan P. Torres}
\affiliation{ICFO—Institut de Ciencies Fotoniques, the Barcelona Institute of Science and Technology, 08860 Castelldefels (Barcelona), Spain}
\affiliation{Department of Signal Theory and Communications, Universitat Politecnica de Catalunya, 08034 Barcelona, Spain}

\author{Roberto de J. León-Montiel}
\affiliation{Instituto de Ciencias Nucleares, Universidad Nacional Aut\'onoma de M\'exico, Apartado Postal 70-543, 04510 Cd. Mx., M\'exico}
\email{roberto.leon@nucleares.unam.mx}

\date{\today}
\begin{abstract}
    We demonstrate a \emph{smart} laser-diffraction analysis technique for particle mixture identification. We retrieve information about the size, geometry, and ratio concentration of two-component heterogeneous particle mixtures with an  efficiency above 92\%. In contrast to commonly-used laser diffraction schemes---in which a large number of detectors is needed---our machine-learning-assisted protocol makes use of a single far-field diffraction pattern, contained within a small angle ($\sim 0.26^{\circ}$) around the light propagation axis. Because of its reliability and ease of implementation, our work may pave the way towards the development of novel smart identification technologies for sample classification and particle contamination monitoring in industrial manufacturing processes.
\end{abstract}

\maketitle
\section{Introduction} 

Particle characterization techniques have long played a fundamental role in many different branches of science and technology. In biology, they assist in schemes for the detection of bacteria \cite{Bacteria} and viruses \cite{virus}. They are important in the pharmaceutical  \cite{shekunov2007, dhamoon2018}, food processing \cite{robins_book,ZHANG2020100698}, and the semiconductor \cite{roy2014} industries. Particularly important are applications aimed at environmental monitoring and protection \cite{tinke2008,Pigments2016}. Some potential applications include the detection of microplastics in marine waters \cite{parrish2019microplastic}, and the characterization of airborne particles, given that their size is strongly correlated with pulmonary toxicity \cite{brown2001size,oberdorster2000acute} leading to respiratory illnesses.

Remarkably, more than 75\% of all materials processed in industry are in particulate form. These particles may be presented in any of the three known phases (solid, liquid or gaseous) and can be divided into three broad groups: natural, industrially processed from natural products, and completely synthetic particles \cite{merkus_book}. In general, one can identify two important reasons for industries routinely employing particle characterization \cite{malvern_book}: better understanding of products and processes, and better control of product quality. While the former allows for the optimization of the manufacturing processes, the latter can evidently translate into a potentially important economic benefit. 

During the past two decades, several light scattering technologies used for particle characterization have matured and even become a key part of industrial production lines \cite{xu2015}. These techniques may be classified into three main categories: static light scattering (SLS), dynamic light scattering (DLS), and scattering tracking analysis (STA).  In the first class, the measured scattering signal results from the light-particle interaction at various spatial locations, whereas in the second and third, the recorded signal results from the monitoring of light-particle interaction as a function of time. While dynamic light scattering techniques  can resolve particles deep in the submicron region \cite{stetefeld2016},  static methods work best in the range of hundreds of $\mu m$ to $mm$ \cite{rawle_book,iso_book}. 

Static light scattering, also known as laser diffraction (LD) analysis has become the most widely used technique for extracting information about the particle size distribution of an unknown sample \cite{blott2004}. This technique is based both on Mie light scattering theory, and on far-field Fraunhofer diffraction.  In LD analysis, the light intensity vs scattering angle is related to the dimensions of the particles participating in the scattering process, with other variables, such as wavelength, kept constant.  Thus, information about particle size is extracted from the angular intensity variation of laser light scattered from a given sample: larger particles scatter light at smaller angles, while smaller particles scatter at wider angles \cite{malvern_book}. It is worth pointing out that while for Fraunhofer diffraction the particle size analysis is somewhat straightforward, the Mie scattering approach requires knowledge of the real and imaginary parts of the sample’s refractive index \cite{merkus_book}.    

Commercial LD instruments have been used extensively in industry due to their high precision and reliability. Important drawbacks include their limited portability and their inability to fully discern among different particle shapes. In particular, given that LD is based on the precise detection at different scattering angles, typical instruments require in the region of 16 to 32 detectors positioned at different angles with respect to the main optical axis (see, for instance, Fig. 4 of Ref. \cite{xu2015}). Unfortunately, increasing the number of detectors does not necessarily lead to a better resolution \cite{rawle_book} and thus, finding the optimum number and location of detectors for a particular application becomes a crucial task. Furthermore, the technique is based upon the assumption that particles, although different in size, are always spherical. This evidently poses a problem if the goal is to identify samples containing particles with different shapes \cite{Chen2015,Cooley2018}.

In this work, we provide the first steps towards ``smart'' laser diffraction analysis. This technique makes use of trained artificial Neural Networks (NN) to identify spatial features of heterogeneous mixtures of microscopic objects. Interestingly, neural networks have been shown to significantly improve the detection of spatial features in out-of-the-lab technologies such as mobile-phone-based microscopes \cite{ozcan2018}. The key point is that NN make use of their self-learning capabilities to enhance the performance of optical systems in terms of reliability and resolution without the need to increase its complexity. This undoubtedly facilitates their deployment in industrial scenarios.

The analysis is performed by monitoring the far-field diffraction pattern produced by laser light impinging on two-dimensional arrays of particles which, for the sake of simplicity and generality, are simulated with the help of a Digital Micromirror Device (DMD). We would like to point out that, although these model particles are not real three-dimensional objects, they exhibit certain advantages \cite{hewitt1980,kang1994}. In particular, given their two-dimensional nature, they reasonably fulfill the theoretical pre-assumptions for Fraunhofer diffraction, and by excluding other possible effects derived from real three-dimensional particles help focus our attention on the advantages of using smart technologies for laser diffraction analysis.

Our proposed technique offers two main advantages over typical LD devices. Firstly, it allows for efficient particle identification by detecting the signal within a small angle ($\sim 0.26^{\circ}$) with respect to the light propagation axis, thus effectively reducing the number of detectors needed for its implementation. Note that this property is shared with recent micro- and nano-particle identification proposals that make use of Machine Learning (ML) algorithms \cite{yevick2014, colloids, volpe2019, Silvania_OE, Valerio}. Secondly, our technique permits the identification of particle shapes in two-component heterogeneous mixtures, resolving not only the shape of the particles that make up the mixture, but also the predominance (or balance) between particle geometries. This feature might be relevant for monitoring particle contamination in industrial manufacturing processes \cite{Silvania_AO}. 

\section{Methods} 

\subsection{Experimental setup and data acquisition}

\begin{figure}[b!]
\centering
\includegraphics[width=8cm]{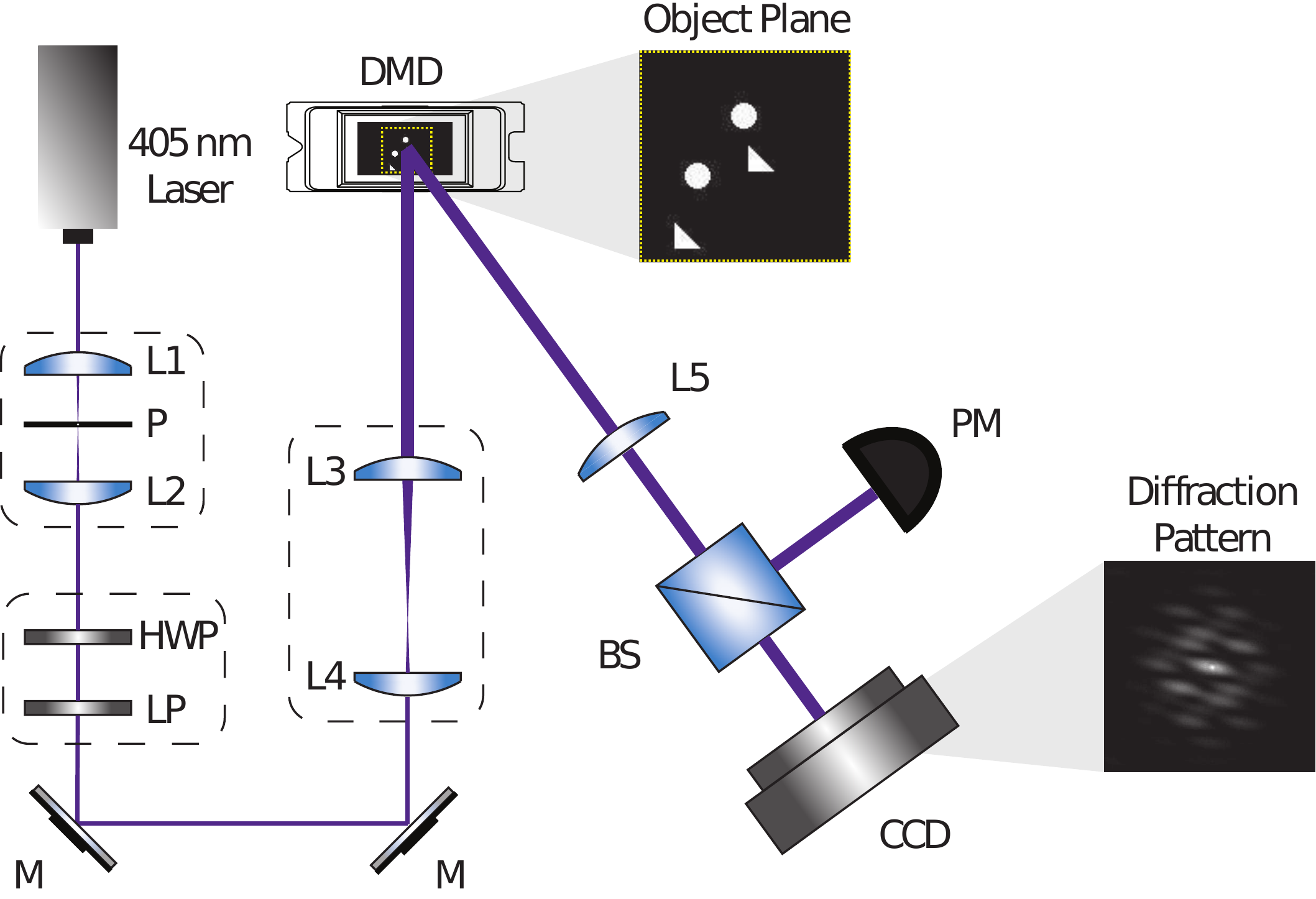}
\caption{A 405 nm laser beam, spatially filtered with two lenses (L1 and L2) and a pinhole (P), is expanded with a telescope system (L3 and L4) to illuminate a Digital Micromirror Device (DMD). Using a Half-wave Plate (HWP) and a Linear Polarizer (LP) the power of the illumination beam can be controlled. The light reflected by the DMD passes through a Fourier transform lens (L5), and the diffraction pattern is collected by a CCD camera at the focal plane of the Fourier lens L5. The power of the signal after L5 was measured using a beam splitter (BS) and a power meter (PM).}
\label{fig:Setup}
\end{figure}

In order to generate a large set of different particle mixture configurations, we create objects of different geometries and sizes using a DMD consisting of a $6.57\, mm \times 3.69\, mm$ chip composed of a grid of square mirrors of $7.63\,\mu m$ per side. The mirrors can be selectively rotated $\pm 12^\circ$ in an "on" or "off" configuration such that when illuminated, the DMD reflects light selectively. 

A collimated laser beam (405 nm wavelength) illuminates the DMD. Each object of the mixture corresponds to a contiguous set of mirrors in the "on" configuration that reflects a part of the beam. The objects are randomly positioned on the DMD plane. Due to the periodicity of the mirrors, a mesh of diffraction order beams is produced. A single diffraction order is selected to be transmitted through a Fourier transform lens, and the diffraction pattern is collected with a CCD camera, as shown in figure \ref{fig:Setup}. We consider mixtures of microscopic particles. The aim is to retrieve information such as their size, geometry, and concentration. These mixtures are analyzed following the steps shown as a flowchart in Fig. \ref{fig:workflow} (a). 

To demonstrate that we can successfully retrieve the sought-after information about the microscopic objects, we carry out two different experiments. The first experiment (Experiment \#1) aims at recognizing sets of microscopic objects that have the same geometry, namely squares, triangles or circles, with different characteristic lengths: 11, 15, 21, or 25 times the DMD mirrors length (7.63 $\mu m$).  The total number of particles varies from one to five. 

The second experiment (Experiment \#2) considers mixtures containing two types (out of the three available geometries) of microscopic objects. In this experiment the size of the particles is kept constant (15 DMD micromirrors), while the total number of objects ranges from 2 to 10. All possible combinations $n_1+n_2=N$ are considered, where $n_1$ and $n_2$ are the number of sources belonging to geometries 1 and 2, respectively. The data-set for each experiment is created by randomly assigning the position of the objects, avoiding any overlap between them, and registering their corresponding far-field diffraction pattern. One hundred diffraction patterns were considered for each category. Given the total number of combinations of size, geometry, and number of objects, Experiment \#1 contains 6000 experimental diffraction patterns, while Experiment \#2 includes 18,900 patterns. 

\begin{figure}[t!]
\centering
\includegraphics[width=\linewidth]{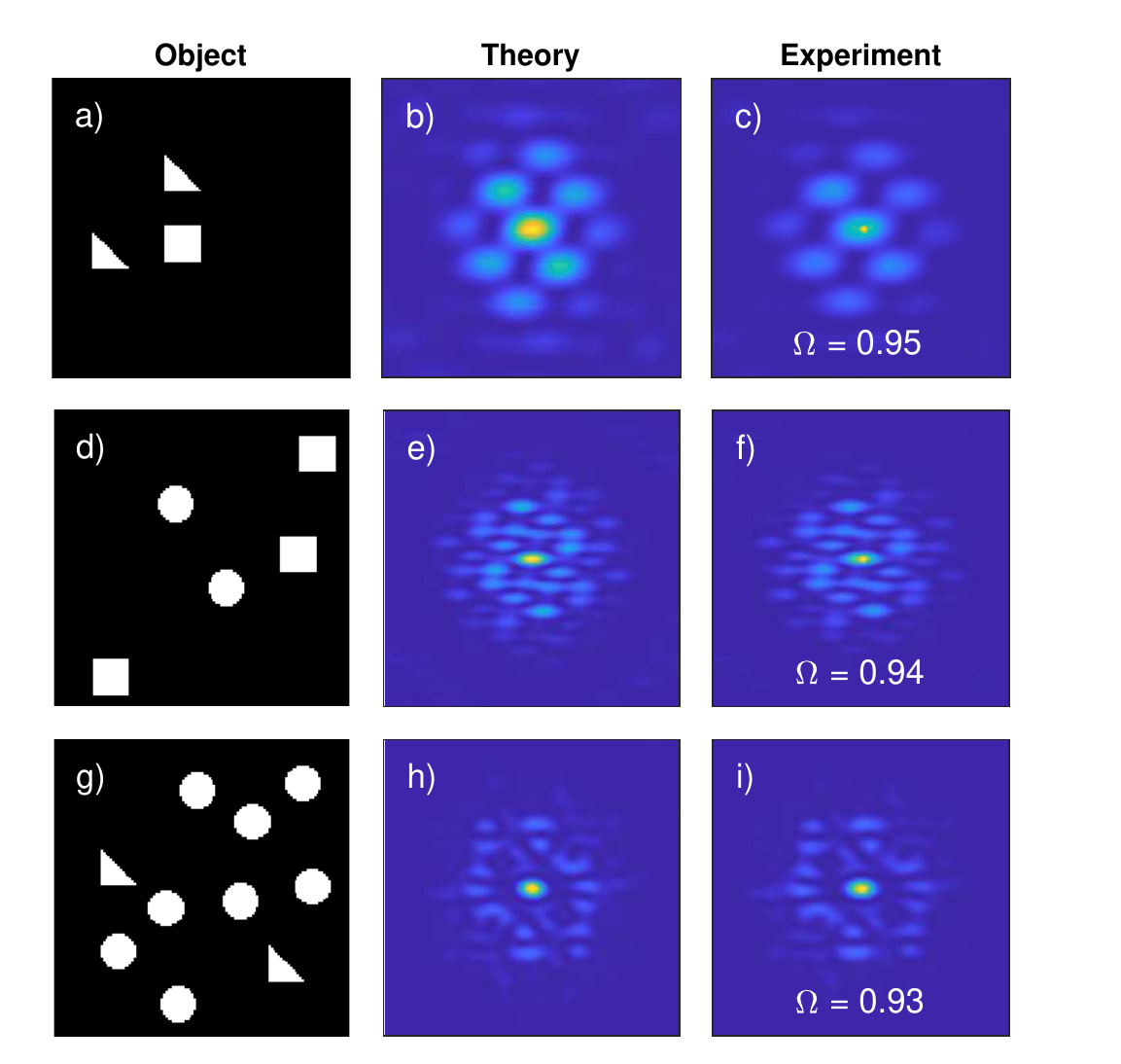}
\caption{Diffraction patterns measured in experiments. {\bf a), d), g)} show examples of the objects generated with the Digital Micromirror Device (DMD). {\bf b), e), h)} are the theoretically predicted diffraction pattern created by the objects depicted in the leftmost column. {\bf c), f), i)} are the experimentally measured diffraction pattern. The overlap parameter $\Omega$ is always larger than $0.9$ in all cases measured.}
\label{fig:Overlap_difraction.}
\end{figure}

Examples of the collected diffraction patterns measured [$I_{E}({\bf x}$)] are shown in Figure \ref{fig:Overlap_difraction.} and compared with the theoretical predictions [$I_{T}({\bf x}$)], where ${\bf x}=(x,y)$ designates the transverse coordinate on the measuring plane. To evaluate the degree of similarity between experiment and theory, we make use of the overlap parameter \cite{armando2018}
\begin{equation}
  \Omega =  \frac{\left[ \int I_{E}^{1/2}({\bf x}) I_{T}^{1/2}({\bf x})\dd {\bf x} \right]^2}{\left[\int I_{E}({\bf x}) \dd {\bf x} \right]\left[\int I_{T}({\bf x}) \dd {\bf x} \right]},
\end{equation} 
where $\Omega=1$ corresponds to a perfect overlap between the theoretical prediction and the experimental measurement.

\subsection{Neural network architecture and processing}

\begin{figure*}[t!]
\centering
\includegraphics[width=18cm]{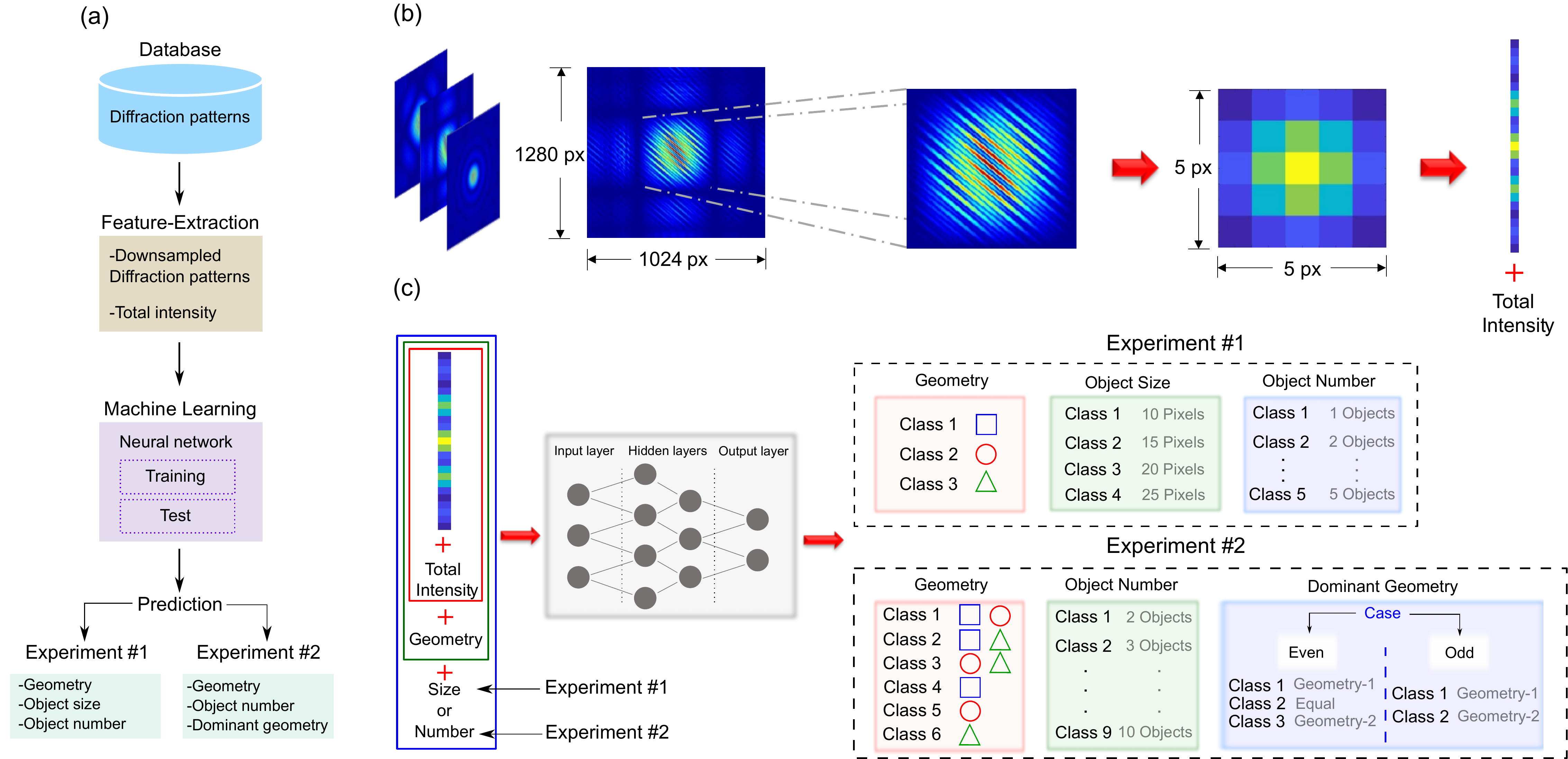}
\caption{Machine-learning-assisted particle mixture identification. (a) Flowchart of the machine learning algorithms used for extracting information from intensity-only measurements of the diffraction patterns. (b) Image down-sampling process. The intensity signal extracted from the CCD camera is presented in false colors for the sake of clarity of the presentation. (c) Flow diagram of the neural networks used in each phase of the experiments described in the main text.}
\label{fig:workflow}
\end{figure*}

All the algorithms used in our protocol are based on multi-layer feed-forward networks \cite{svozil1997introduction}. Hidden layers feature neurons that perform operations on the data using synaptic weights and a nonlinear activation function, the so-called sigmoid function. The output layer comprises softmax neurons that provide a probability distribution over predicted output classes \cite{Goodfellow2016,bishop2006pattern}. So as to build accurate and reliable neural networks, it is necessary to include relevant features that capture the encoded information in the diffraction patterns. Figure \ref{fig:workflow}(b) shows the image pre-processing method carried out to build the feature vector. We first crop the diffraction pattern to a $400 \times 400$ pixel image, retaining only the central portion of the monochromatic high-resolution original images ($1280 \times 1024$ pixels, normalized to 8 bits) obtained with a CCD camera (Thorlabs DCU224C). After this step, in order to reduce the data dimensions, we perform a down-sampling process that averages small clusters of 80 by 80 pixels, resulting in a 5-by-5 pixel image. Finally, we rearrange the resulting intensity distribution as a column vector, where the total measured intensity is included as the 26th element of the feature vector $V_{1}$, depicted by the red rectangle in Fig. \ref{fig:workflow}(c). 

Our neural networks undergo two stages, training and testing. We train the classification networks by using the scaled conjugate gradient back-propagation algorithm \cite{moller1993scaled}, while the performance is evaluated with the cross-entropy \cite{shore1981properties,de2005tutorial}.  We devoted 70\% of the dataset to training, 15\% to validation, and 15\% to testing, as is standard in ML protocols \cite{shebani2018prediction, you2020identification}. It is worth mentioning that the testing data was excluded from the training phase. It provides an unbiased evaluation of the algorithm's overall accuracy.  A limit of 1000 epochs was set for each network training stage. After training, our networks can make predictions of the geometry, size, and number of microscopic objects in a given sample using as input the diffraction pattern and the total intensity, as shown in Figure \ref{fig:workflow}(c). In what follows, we provide a thorough description of the steps followed in each experiment.

In Experiment \#1, we implement three neural networks connected in series, each one of them performing a specific prediction of the features of the initial field. The first neural network  identifies the geometry of the objects; it is trained by using a concatenation of the total intensity (signal power) and the down-sampled representation of the diffraction pattern, i.e., the feature vector $V_{1}$ (see Fig. \ref{fig:workflow}(b). The second network, identifying the object size, makes use of the prediction of the first network to create the feature vector $V_{2}=V_{1}+\text{Geometry}$, whereas the third network extracts the object number from the feature vector $V_{3} = V_{2}+\text{Size}$. Figure \ref{fig:workflow}(c) summarizes the structure of the neural networks and the predicted classes. 
The core feature-vector $V_{1}$ is initially introduced in the network that identifies the geometry. Then, its output works as input for the second network, in conjunction with the core feature-vector. Once the object size has been determined, the third neural network predicts the number of objects using as input the core feature-vector, as well as the outputs of the first and second networks.

In Experiment \#2, we follow a similar strategy. We first implement a neural network to determine the combined-geometry class---i.e., the two shapes of the objects which make up the mixture---by using the feature-vector comprising the down-sampled diffraction pattern and the total measured intensity. With the geometry-class identified and the core feature-vector, we then determine the total number of objects. Finally, by making use of the core feature-vector, as well as the outputs of the first and second networks, we predict the dominant geometry. Note that the last network has been divided into two cases, even and odd number of objects. This is due to the fact that when the number of objects is even, we need to define three output classes, whereas for an odd number of objects, there are only two relevant classes. Also note that the class labeled as Geometry-1 in Fig. \ref{fig:workflow} always refers to the first object-shape in each geometry class.

\section{Results and discussion} 

\begin{figure*}[t!]
\centering
\includegraphics[width=18cm]{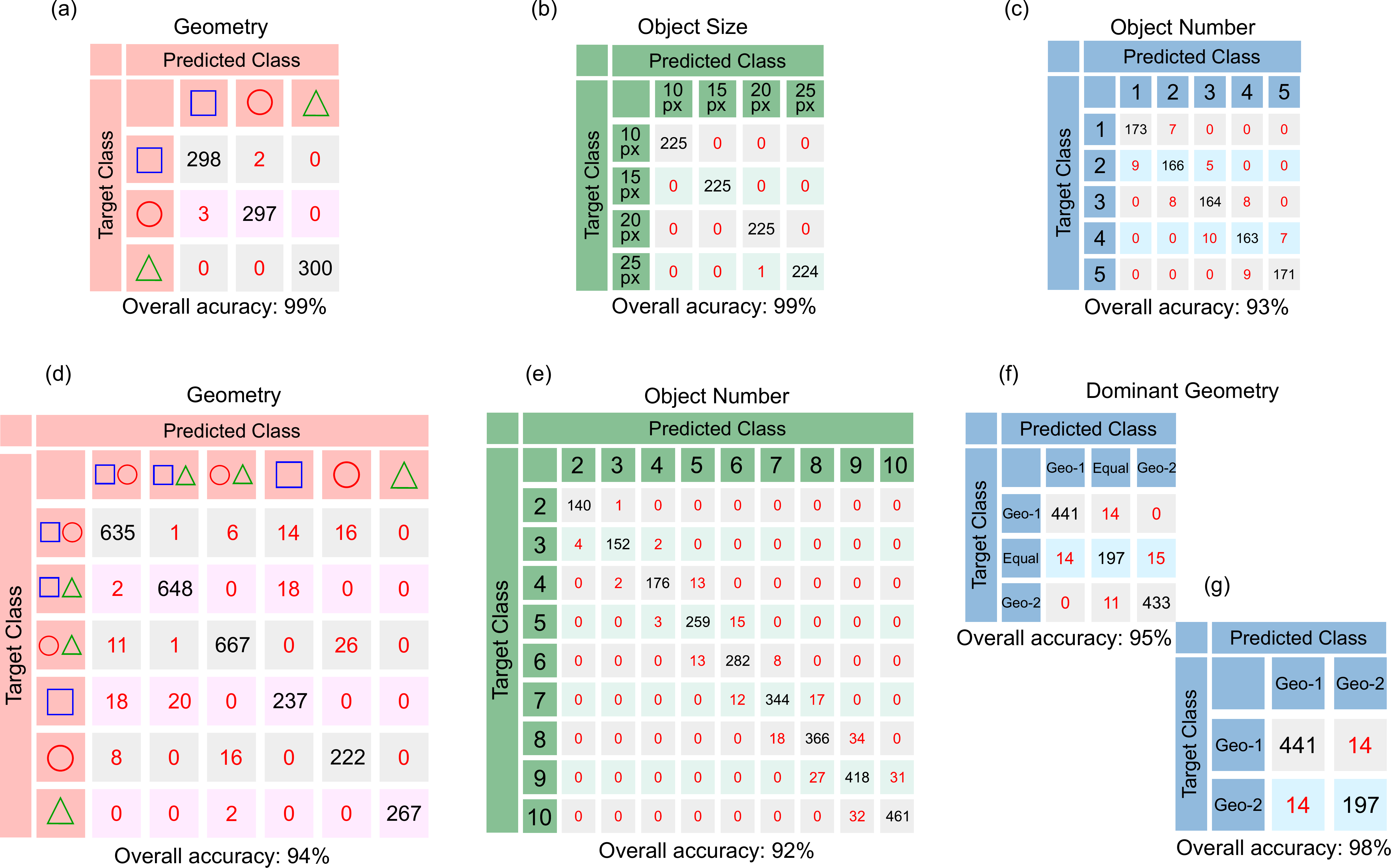}
\caption{Confusion matrices that summarize the performance of machine-learning algorithms for particle mixture identification. The top row shows the confusion matrices containing information about the correct and incorrect predictions for (a) geometry, (b) object size, and (c) number of objects of the Experiment \#1.  The bottom row presents the confusion matrices for (d) geometry, (e) number of objects, and (f)-(g) dominant geometry of the Experiment \#2. (f) and (g) matrices correspond to the case in which the number of objects is even and odd, respectively. The diagonal elements of the matrices represent successful recognition, i.e., true-positives and true-negatives, whereas off-diagonal elements represent failed attempts, false-negatives, and false-positives.}
\label{fig:matrix1}
\end{figure*}

We have performed a blind test of all NN on the remaining 15\% of the collected data left of the training. We found an overall >90\% identification efficiency in every stage of the experiments. Table \ref{tab:results} summarizes the results of each experiment (see Fig. \ref{fig:matrix1} for details on the network success rate for each task), including the overall accuracy, number of hidden layers, and number of neurons in each layer.

In Experiment \#1, the first NN was able to identify the geometry of the particles with an accuracy of 99\% using a single hidden layer with 5 neurons. The second network for the particle size showed the same performance with the same architecture. For the case of the particle number, the architecture of the NN was different, namely two hidden layers with 20 and 5 neurons, respectively, resulting in a 93\% identification accuracy. 

Experiment \#2 required two hidden layers per NN; the geometry of the particles was retrieved with 94\% accuracy using 30 and 20 neurons, respectively; whilst for the total number of particles, 80 and 50 neurons were needed for an identification of 92\%. Finally, two layers with 30 and 20 neurons were required to determine the dominant geometry obtaining 95\% accuracy for the even total number of particles and 98\% for the odd number of particles.

It is worth mentioning that although the data preparation and processing can be considerably time-consuming, once the network has been trained our technique is in fact  extremely fast.  Our algorithm can process newly-acquired data (prepared in the same format as used for training) in timescales of milliseconds.

\section{Conclusions}

We have demonstrated an optical technique for particle mixture identification, with potential applications in research and industry, based on machine-learning-assisted laser diffraction analysis. The technique proposed facilitates a fast and accurate identification of the particle's geometry and size, as well as the dominance of one substance over another in a heterogeneous mixture of particles. It is worth pointing out that equivalent machine learning and deep learning algorithms have been used to improve optical microscopy \cite{DLM_survey, DLmic, machine_vision,ozcan2018}. Interestingly, by using neural networks as classifiers or feature extractors, in some cases trained with synthetic data and tested in real measurements, this type of techniques have shown to be effective for impurity recognition in semiconductors \cite{Wafer, chem_comp}.

In our work, by making use of a digital micromirror device, we have simulated mixtures of particles of different sizes and geometries with a different total number of particles. By using the resulting far-field diffraction pattern, our neural network algorithm was able to extract the spatial features of the mixtures. Relying on a total of 24,900 diffraction patterns and a 70/15/15 ratio for training, validation and testing data, respectively, the identification performance was always above 90\%. Because of its reliability and ease of implementation, our technique may be of great importance for different scientific and technological disciplines, as it establishes a new route towards the development of novel smart identification devices for sample classification and particle contamination monitoring.

\begin{table*}[t!]
 \centering \caption{\bf Overall accuracy, number of hidden layers, and number of neurons in each layer for the neural networks implemented in the described experiments.}
 
 \begin{tabular}{clccc}
 \hline
Experiment & Neural Network & Accuracy & Number of Hidden Layers & Number of Neurons by Layer\\
 \hline
\multirow{3}{*}{1}&Geometry & 99\% & 1& 5\\
                  &Object Size & 99\% & 1& 5\\
                  &Object Number & 93\% & 2& Layer 1= 20; Layer 2= 5\\
 \hline
 \multirow{4}{*}{2}&Geometry& 94\% & 2& Layer 1= 30; Layer 2= 20 \\
                  &Object Number & 92\% & 2& Layer 1= 80; Layer 2= 50 \\
                  &Dominant geometry (even) & 95\% & 2& Layer 1= 30; Layer 2= 20 \\
                  & Dominant geometry (odd) & 98\% & 2& Layer 1= 30; Layer 2= 20 \\
\hline
 \end{tabular}
   \label{tab:results}
 \end{table*}

\section*{Acknowledgments}
This work was partially funded through the EMPIR project
17FUN01-BeCOMe. The EMPIR initiative is co-funded by the European Union Horizon 2020 research and innovation
programme and the EMPIR participating States.
A. V. thanks the financial support from PREBIST that has  received  funding  from  the  European Union’s Horizon   2020  research and innovation   programme under the Marie Sklodowska-Curie grant agreement No 754558. M.A.Q.J. and R.J.L.M. thankfully acknowledge financial support by CONACyT under the project CB-2016-01/284372, and by DGAPA-UNAM under the project UNAM-PAPIIT IN102920. M.A.Q.J. thanks DGAPA-UNAM posdoctoral fellowship number 1595/2020.   A.U. acknowledges support from DGAPA-UNAM through grant UNAM-PAPIIT IN103521, and CONACyT through "Ciencia de Frontera" grant 217559.

\newpage
\bibliography{main}
\end{document}